# Readout of field induced magnetic anisotropy in a magnetoactive elastomer

Andrii V. Bodnaruk[1], Alexander Brunhuber[2], Viktor M. Kalita[1,3], Mykola M. Kulyk[1], Andrei A. Snarskii[3,4], Albert F. Lozenko[1], Sergey M. Ryabchenko[1], Mikhail Shamonin[2*]

[1]*Institute of Physics, NAS of Ukraine, Prospekt Nauky 46, 03028 Kiev, Ukraine*
[2]*East Bavarian Centre for Intelligent Materials (EBACIM), Ostbayerische Technische Hochschule (OTH) Regensburg, Prüfeninger Strasse 58, 93049 Regensburg, Germany*
[3] *National Technical University of Ukraine "Igor Sikorsky Kyiv Polytechnic Institute", Prospekt Peremohy 37, 03056 Kiev, Ukraine*
[4]*Institute for Information Recording, NAS of Ukraine, Shpaka Street 2, 03113 Kiev, Ukraine*

It is shown that in external magnetic fields, a uniaxial magnetic anisotropy comes into being in a magnetoactive elastomer (MAE). The magnitude of the induced uniaxial anisotropy grows with the increasing external magnetic field. The filler particles are immobilized in the matrix if the MAE sample is cooled below 220 K, where the anisotropy can be read out. The cooling of the sample is considered as an alternative methodological approach to the experimental investigation of the magnetized state of MAEs. The appearance of magnetic anisotropy in MAE is associated with restructuring of the filler during magnetization, which leads to an additional effective field felt by the magnetization. It is found that the magnitude of the effective magnetic anisotropy constant of the MAE is approximately two times larger than its effective shear modulus in the absence of magnetic field. It is proposed that the experimentally observed large (about 40) ratio of the magnetic anisotropy constant of the filler to the shear modulus of the matrix deserves attention for the explanation of magnetic and magnetoelastic properties of MAEs. It may lead to additional rigidity of the elastic subsystem increasing the shear modulus of the composite material through the magnetomechanical coupling.

**Keywords:** Magnetoactive elastomer; magnetic properties; magnetorheological elastomer; magnetomechanical coupling; experimental methodology

---

[*] Corresponding author. E-Mail: mikhail.chamonine@oth-regensburg.de

## 1. Introduction

Magnetoactive elastomers (MAEs) are composite materials where micrometer sized ferromagnetic inclusions are embedded into a compliant elastomer matrix[1-6]. The ability of these ferromagnetic inclusions (filler particles) to be displaced and/or rotated due to the presence of an external magnetic field and interactions between magnetized particles distinguishes these composite materials from their conventional counterparts[7-12]. At room temperature, the elasticity of the matrix forces the soft magnetic filler particles of the sample towards their initial positions when the magnetic field vanishes. The properties of the composite material with soft magnetic inclusions are largely recovered in the "quiescent" state[13], despite the large displacements of particles comparable with their dimensions when the samples are deformed by magnetic fields[7]. It has been recently suggested that at sufficiently low temperatures the iron particles can be immobilized in a polydimethylsiloxane (PDMS) matrix[14] because the matrix becomes rigid[15]. This implies that the filler particles are blocked for displacement and do not have the ability to rearrange during magnetization at low temperatures. The possibility of displacement and rotation of particles in external magnetic fields is commonly believed to be the origin of several unique properties of MAEs, namely its anomalous magnetostriction[7,9,16-21], giant magnetic field dependence of elastic moduli[9,22-26], and strong variation of electric properties in magnetic fields[27-36].

Obviously, these MAE properties are related to their magnetization. Several experiments on magnetization of MAEs have been reported in the literature[7,37-42]. Hitherto, the research has concentrated on investigation of the magnetic hysteresis in MAEs filled with soft magnetic particles, which is characterized by the absence of both the coercive force and the remanent magnetization[40,42]. In the context of memristor research, such a behavior is called "pinched" hysteresis loop[43,44]. It is assumed that in MAEs such a hysteresis is associated with the mobility of filler particles inside the polymer matrix[1,42] and the corresponding assemblage of magnetized particles into elongated ("linear") chain-like aggregates[45] during the ascending magnetization branch and disintegration of these "chains" under the action of the elastic forces when the magnetic field is decreased[40]. It is further proposed[46] that the decay scenario does not coincide with the scenario of chain formation, which can lead to hysteresis behavior. This simplified physical picture for high concentrations of magnetizable particles has been recently questioned in Ref. 47, where numerical simulations showed that formation of elongated structures may become impossible due to purely geometrical constraints.

The physical intuition suggests that the external magnetic field should induce the anisotropy of initially isotropic compliant MAEs in a sense that their physical (e.g. magnetic) properties will be different in the directions along and perpendicular to the direction of the magnetic field. Magnetic anisotropy can arise due to the influence of the magnetoelastic field when the sample experiences magnetostriction in a magnetic field[48,49]. It has been experimentally (see, e.g. Fig. 6 in Ref. 7) and theoretically verified that the physical properties of structured MAEs, where the particles have been aligned along the magnetic field lines during the cross-linking procedure, are anisotropic[50-53]. We consider the case of a restricted sample with an initially unstructured material. If an MRE sample is synthesized in the absence of a magnetic field, it has an isotropic structure[54,55]. However, in the initially isotropic compliant MAE the restructuring of the filler will "follow" the magnetic field and therefore is not easy to detect. We are not aware of any previous publications where the magnetic anisotropy of magnetized MAEs was measured.

In the present paper, we show that an MAE material, which is isotropic in the absence of an applied magnetic field, in sufficiently large magnetic fields acquires a magnetic anisotropy, whose constant exceeds the elastic constant of the matrix (these quantities have the same dimension). The composite material comprises soft-magnetic, randomly distributed inclusions. The sample dimensions are fixed. Therefore, there is no overall magnetostriction, but the embedded inclusions can move inside the sample. It has to be expected that the magnitude of induced magnetic anisotropy depends on the magnetization of the sample. The appearance of the magnetic anisotropy is associated with restructuring of the filler, i.e. with the change of the positions and/or orientations of the particles in the sample under the action of internal magnetic forces from inter-particle interactions.

If the filler particles inside the MAE can change their position and/or orientation in magnetic fields, it is necessary to fix them in positions corresponding to the magnetized state for measuring the magnetic anisotropy. It is known[15] that as the temperature is lowered below a characteristic temperature, the elastic moduli of the PDMS matrix can increase by several orders of magnitude. We have used this property of the matrix in our previous work [14] to prove that the particles are displaced when the MAE is magnetized at room temperature. In the present paper, we will use this distinctive attribute of the matrix for investigating the properties of the magnetized state of the MAE, namely the occurrence of anisotropy in a magnetized MAE. In this way, we utilize the "solidification" of the matrix for fixing (blocking) the positions of filler particles in the magnetized state. In this case, the structure of the positions of the particles after freezing will be the same as it was in the magnetized MAE sample at room temperature, and thereby we find that freezing allows one to investigate the properties of MAEs magnetized at room temperature. Thus, the freezing of the sample should be considered as a new experimental technique for studying the magnetized state of MAE materials.

Therefore, the purpose of our work is to investigate the magnetic anisotropy that arises in the magnetized MAE. It will be shown that this magnetic-field-induced anisotropy is uniaxial, where the easy axis of magnetization is directed along the magnetic field. It will be demonstrated that the process of formation of magnetic anisotropy is nonlinear in a sense that the anisotropy constant nonlinearly depends on the magnitude of the external field. Note that this magnetic anisotropy is not a consequence of cooling of the sample. It appears when the MAE sample is magnetized at room temperature, and cooling of the sample is a methodological feature of the experiment. The magnetic anisotropy field is an additional field that arises in a magnetized MAE. It also follows from our work that the formation of this additional effective field is a consequence of a nonlinear self-consistent process related to restructuring of the filler.

## 2. Experimental

A sample comprising carbonyl iron particles with an average particle size of 4.5 μm (type SQ, BASF SE Carbonyl Iron Powder & Metal Systems, Ludwigshafen, Germany) embedded into a PDMS-based elastomer matrix was investigated. This is the same composite material as described in Ref. 14. The base polymer VS 100000 (vinyl-functional polydimethylsiloxane (PDMS)) for addition-curing silicones, the chain extenders Modifier 715 (SiH-terminated

PDMS), the reactive diluent polymer MV 2000 (monovinyl functional PDMS), the crosslinker 210 (dimethyl siloxane-methyl hydrogen siloxane copolymer), the Pt-catalyst 510 and the inhibitor DVS were provided by Evonik Hanse GmbH, Geesthacht, Germany. The silicone oil WACKER® AK 10 (linear, non-reactive PDMS) was purchased from Wacker Chemie AG, Burghausen, Germany.

The polymer VS 100000, the polymer MV 2000, the modifier 715 and the silicone oil AK 10 were put together and blended with an electric mixer (Roti®-Speed-stirrer, Carl Roth GmbH, Germany) to form an initial compound. In the next step, the initial compound was mixed together with the CIP particles and the crosslinker 210. The crosslinking reaction was activated by the Pt-Catalyst 510. For the control of the Pt-catalyst's activity, the inhibitor DVS was used, the recommended dosage according to the literature is between 0,01 and 0,5 %.. The MAE samples were pre-cured in the universal oven Memmert UF30 (Memmert GmbH, Schwabach, Germany) at 353 K for 1 hour and then post-cured at 333 K for 24 hours with air circulation.

From the literature[56,57], it is known that the shape of the filler particles is close to spherical, although, to the best of our knowledge, the statistical analysis of the deviations from the shape of the sphere was not carried out. The distribution of the particle size in the iron powder was investigated in Ref. 58. The mass fraction of iron particles in the sample is 70 wt.%, and the corresponding relative occupation of the sample's volume by the filler is approximately 22%. The sample was placed in a rigid cuvette at room temperature in the absence of a magnetic field and filled it completely. The cuvette is a cylinder with the height of 2 mm and the diameter of 2.5 mm. Thus, the sample in the cuvette is forced to retain its shape and size when the magnetic field is applied. Obviously, under these conditions, internal stress could arise in the sample because of the obstruction of its magnetostriction by the cuvette walls. As a whole, the described sample should display a magnetic anisotropy associated with its demagnetizing tensor due to the constant non-spherical shape of the sample.

The sample consisted of ferromagnetic particles distributed in a non-magnetic matrix. The distribution of particles can be considered uniform. There were no indications of sedimentation of particles and no magnetic field was applied during crosslinking. During magnetization of the sample, filler particles could be displaced and rotated under the influence of magnetic interactions with the external field and between the particles, against the resulting elastic forces in the deformed matrix[7-11]. It will be shown below that those changes of the positions of the particles and their rotations inside the sample (within the accuracy of measurements) do not influence the effective demagnetization factor of the sample under conditions of rigid limitation of the sample shape. The experimentally determined effective demagnetizing factor is independent of external magnetic field.

The restriction of the external shape of the sample by a rigid cuvette was also motivated by the fact that the vibrating sample magnetometer (LDJ-9500, LDJ Electronics, Troy, MI 48099, USA) used for magnetostatic measurements creates a working acceleration of $25g_n$, where $g_n$ is the standard gravity. Such acceleration could cause dynamic deformations of the sample shape, which would distort the results of measurements. The rigid container prevented this source of errors. The magnetometer measures the component of the magnetization of the sample along the direction of the magnetic field created by the device's electromagnet.

## 3. Magnetization at room temperature

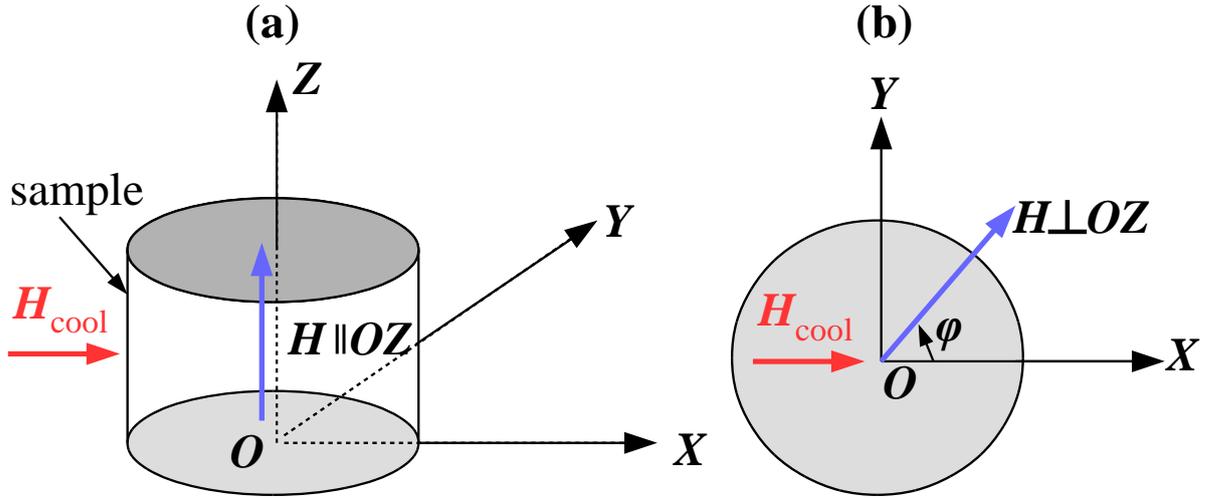

**Figure 1.** Experimental geometry and the coordinate system. (a) Orientation of the sample in the field $H_{cool}$ and the measurement field $H\|OZ$. (b) Cross-section of the sample and the measurement field $H\perp OZ$.

Figure 1 shows the experimental geometry. The basic property of the magnetostatic measurements of this sample in the magnetic field range between -10 and +10 kOe is that the $m(H)$-dependences were completely reproducible after repeated recording at room temperature. $m$ denotes the magnetization of the composite material and $H$ is the external magnetic field strength. There were only minor differences in the reproduced curves from the initial magnetization curve of the demagnetized sample. This unambiguously indicates that all the displacements (or possible rotations) of the particles that could have occurred during the magnetization reversal of the sample were reversible in a sense that the elasticity of the matrix restores the initial arrangement of the particles in the absence of magnetic field.

Figure 2 presents the magnetization curves. Curve 1 is measured when the magnetic field is directed perpendicular to the axis of the cylinder ($H\perp OZ$). Curve 2 is obtained for the case when the magnetic field is directed along the axis $OZ$ of the cylinder ($H\|OZ$). The magnetization dependences for all directions of the external field in the sample plane were all the same (i.e., like curve 1). The differences between curves 1 and 2 can be explained by the effect of the demagnetizing magnetic field, which is related to the shape of the sample. Curves 3 in Fig. 1 correspond to the recalculation of curves 1 and 2 with respect to the in the internal magnetic field $H_{int} = H + H_d$, where $H_d$ is the vector of the demagnetizing field, which depends on the components of the demagnetizing tensor (shape factor) of the sample and its average magnetization[59]. The dependencies $m(H_{int})$ should be the same for magnetization directed along the axis of the sample cylinder and perpendicular to it. To achieve the desired equality for re-calculation of curves 1 and 2 into a single curve 3 ($m(H_{int})$ is the property of the composite material) in Fig. 2, we experimentally found the values of the components of the demagnetization tensor of the sample under study: the component $N_\|$ along the $OZ$-axis is equal to 5.60 and the component $N_\perp$ perpendicular to the $OZ$-axis is equal to 3.48. The equality $N_\| + 2N_\perp = 4\pi$ is fulfilled. Note that two curves 3 originating from curves 1 and 2 are indistinguishable on the scale of Fig. 1.

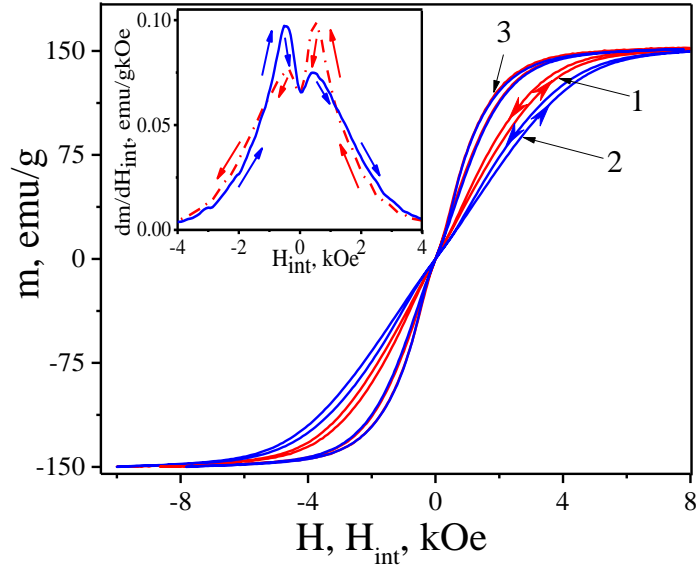

**Figure 2.** Magnetization curves of the MAE sample. Curve 1 is measured in magnetic field directed perpendicular to *OZ*: ***H***⊥*OZ*, and curve 2 is measured in magnetic field directed along the axis of the cylinder: ***H***∥*OZ*. The number 3 designates the overlapping curves derived from curves 1 and 2, re-calculated to the dependence of magnetization *m* on the internal magnetic field $H_{int}$ as described in the text. The inset shows the field dependence of $dm/dH_{int}$ on $H_{int}$. The solid curve presents the course of the differential magnetic susceptibility ($dm/dH_{int}$) for increasing magnetic field. The dotted line shows the differential magnetic susceptibility for decreasing magnetic field. The arrows designate the direction of field change.

Curves 1, 2 in Fig. 2 demonstrate a peculiar hysteresis behavior (sometimes denoted as "pinched" hysteresis[43,44]). Significant hysteresis arises at relatively high fields (> 1 kOe) and it is absent in the vicinity of the zero field, where the magnetization of the particles vanishes, and for sufficiently large fields (> 5 kOe), where the magnetization saturates. However, it is known that carbonyl iron microparticles do not possess significant magnetic hysteresis at room temperature (see e.g. Fig. 8 in Ref. [60]).

Figure 2 demonstrates the possibility of re-calculating the hysteresis curves 1 and 2 into a single curve 3, possessing a hysteresis behavior as well, by using the components of the demagnetization factor tensor, which do not depend on the magnitude of the external magnetic field. Figure 2 also shows the field dependence for the differential magnetic susceptibility $dm/dH_{int}$ of the MAE. The differential magnetic susceptibility has maxima in its dependence of internal magnetic field. The appearance of these maxima can be explained by the elastic resistance of the matrix to the displacement of particles under the action of magnetic forces[14].

Despite the mobility of MAE particles, the sample can be considered as a medium with an effective (homogeneous) magnetization, and provided that the dimensions and shape of the sample do not change, it can be characterized by a demagnetizing tensor that does not depend on the applied field.

## 4. Field dependences of magnetization at low temperatures

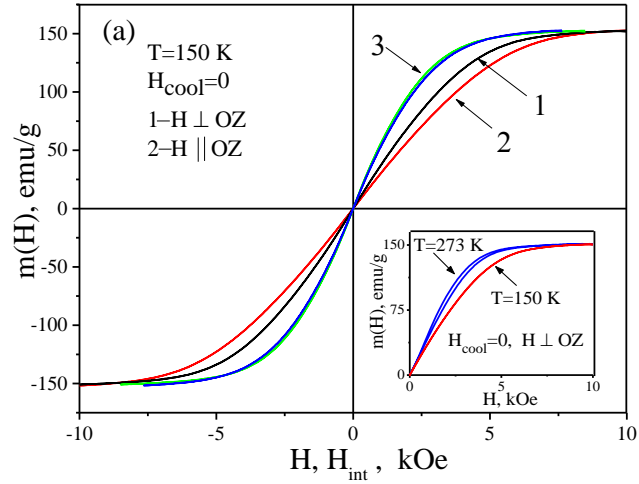

**Figure 3.** Field dependences of the magnetization at $T = 150$ K, cooled in the absence of a magnetic field ($H_{cool} = 0$). Curve 1 is obtained for magnetic fields $H \perp OZ$, curve 2 is measured for magnetic fields $H \| OZ$. Curves 3 ($m(H_{int})$) depict the modified curves 1 and 2, recalculated as a function of the internal magnetic field $H_{int}$. The inset shows the magnetization curves for $H \perp OZ$ at $T = 273$ K and at $T = 150$ K.

Field behavior of the magnetization was also investigated at low temperatures (below 220 K). It turned out that hysteresis of magnetization is not observed (see Fig. 3). The absence of hysteresis at low temperatures is explained by the absence of intrinsic hysteresis of the magnetization of soft-magnetic filler particles[61]. The magnetization hysteresis at higher temperatures (above 230 K) is related to the displacement of particles under the influence of magnetic forces[14]. At a low temperature, when the matrix becomes rigid, the particles loose ability to move within the sample. If there is no hysteresis in a ferromagnetic system at low temperature, then its appearance at a high temperature (Fig. 2) is at first glance counterintuitive.

If the sample is cooled in the absence of an external magnetic field ($H_{cool} = 0$), the magnetization at low temperatures has only anisotropy associated with the effect of the demagnetization factor of the sample. Figure 3 shows the magnetization curves for the external field $H \perp OZ$ (curve 1) and $H \| OZ$ (curve 2). Curves 3 denote the re-calculated dependence of the magnetization on the internal field, where the same components of the demagnetizing factor as in Fig. 2, found at room temperature, were used. It can be seen that the magnetizations measured for different directions of the external field will become practically identical after re-calculation to the internal field. Therefore, at low temperatures, when the particles become immobile in a magnetic field, the values of the components of the demagnetization factor remain the same as at room temperature. If $H_{cool} = 0$, the magnetization at all temperatures is isotropic in the plane perpendicular to the axis of the cylinder.

The inset in Fig. 3 shows the field dependences for the magnetization at $T = 273$ K and at $T = 150$ K. Magnetization at low temperatures has a lower differential susceptibility. These dependences illustrate the differences in the magnetization of a composite material with filler particles immobile and mobile in magnetic fields. As it will be shown below, the origin of these differences is related to the magnetic anisotropy of the MAE induced by an external magnetic field. Such anisotropy, where an easy axis directed along the magnetic field, promotes magnetization of the MAE and increases the magnetic susceptibility.

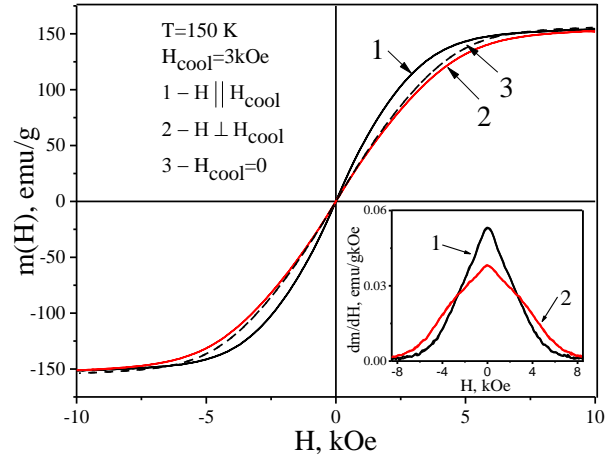

**Figure 4**. Field dependences of the magnetization $m(H)$ at $T = 150$ K, frozen in a magnetic field $H_{cool} \perp OZ$ with $H_{cool} = 3$ kOe. The curve 1 is obtained for $H \| H_{cool}$, the curve 2 is measured for $H \perp H_{cool}$; in both cases $H \perp OZ$. The dashed curve 3 is obtained at $H \perp OZ$ with $H_{cool} = 0$. The inset displays the field dependence of the magnetic susceptibility $dm/dH$ for $H \| H_{cool}$ (curve 1) and for $H \perp H_{cool}$ with $H \perp OZ$ (curve 2).

If the sample is magnetized at room temperature by the field $H_{cool} \neq 0$, which is directed as $H_{cool} \perp OZ$, and then frozen in this field, the isotropy of its magnetic properties in the plane perpendicular to the *OZ*- axis is destroyed. Under the influence of $H_{cool} \perp OZ$, the magnetization anisotropy in the plane perpendicular to the *OZ*-axis comes into play. Figure 4 compares magnetization curves of a sample frozen in the field $H_{cool} = 3$ kOe, which is directed perpendicular to the *OZ*-axis. The curve 1 is obtained for $H \| H_{cool}$, and the curve 2 is obtained for $H \perp H_{cool}$ and $H \perp OZ$. The sample is magnetized easier if $H \| H_{cool}$. The magnetization turns out to be more difficult in the direction $H \perp H_{cool}$, but its dependence $m(H)$ differs only little from the curve 3 obtained at $H_{cool} = 0$.

Thus, when the MAE is magnetized at room temperature by the field $H_{cool}$, the magnetic anisotropy emerges in the sample. The easy axis is directed along the field direction $H_{cool}$. Note that the anisotropy axis is not a vector quantity, like $H$.

The appearance of the anisotropy induced by the magnetic field in the MAE is also confirmed by the field dependence of the differential magnetic susceptibility $dm/dH$, the value of which at $H \to 0$ for the case $\boldsymbol{H}||\boldsymbol{H}_{\text{cool}}$ (curve 1 in the inset of Fig. 4) exceeds the same physical quantity at $H \to 0$ for $\boldsymbol{H} \perp \boldsymbol{H}_{\text{cool}}$ with $\boldsymbol{H} \perp OZ$ (curve 2 in the inset of Fig. 4 ).

Figure 4 can be also seen as an indication that "liner chains", where magnetized particles almost stick to each other, are not formed along the magnetic field. Previously, similar conclusion was derived in Ref. 51 from the ferromagnetic resonance experiments on PDMS matrix filled with $Fe_3O_4$ nanoparticles. It is clear that the demagnetizing factor of such structures in the direction perpendicular to $\boldsymbol{H}_{\text{cool}}$ must be larger than that observed in our sample, and the curve 2 should go significantly lower[62-64] than the curve 3 in Fig. 4.

## 5. Angular dependences of magnetization

Figure 5 shows the family of field dependences at $T = 150$ K for the magnetization of the sample, frozen in the field $H_{\text{cool}} = 5$ kOe higher than it was shown in Fig. 3 and $\boldsymbol{H}_{\text{cool}} \perp OZ$. The dependences have been measured in the field $\boldsymbol{H} \perp OZ$, but which direction deviated from $\boldsymbol{H}_{\text{cool}}$ by an angle $\varphi$ (see Figure 1). It can be observed that the magnetic anisotropy induced by the field $H_{\text{cool}} = 5$ kOe is larger than the anisotropy caused by the field $H_{\text{cool}} = 3$ kOe (see Figure 4). Moreover, the magnetization curve for $\boldsymbol{H} \perp \boldsymbol{H}_{\text{cool}}$ obtained at $H_{\text{cool}} = 5$ kOe differs more from the curve at $H_{\text{cool}} = 0$, than a similar curve obtained at $H_{\text{cool}} = 3$ kOe in Fig. 4.

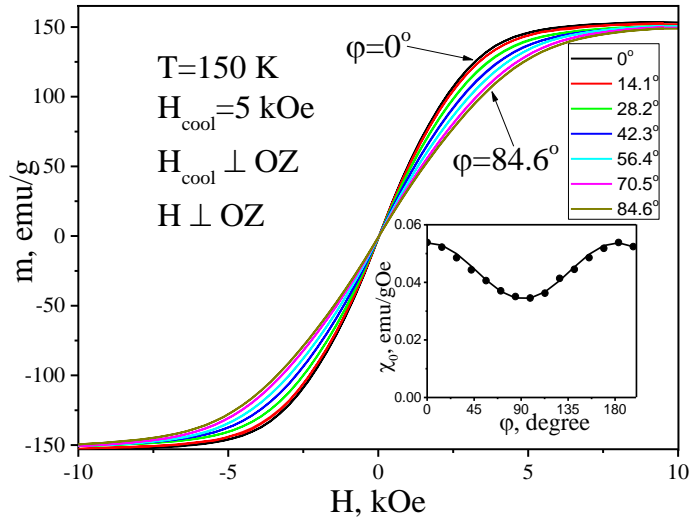

**Figure 5.** Field dependences of the magnetization of the sample $m(H)$ at $T = 150$ K frozen at $H_{\text{cool}} = 5$ kOe and $\boldsymbol{H}_{\text{cool}} \perp OZ$ for magnetic fields inclined at an angle $\varphi$ to the field $\boldsymbol{H}_{\text{cool}}$ in the plane of the sample ($\boldsymbol{H} \perp OZ$). The inset shows the angular dependence of the initial magnetic susceptibility $\chi_0(\varphi)$. In the inset, the black circles denote the experimental values, while the solid curve is the fitted function (1).

The inset in Fig. 5 shows the angular dependence of the magnetic susceptibility $\chi_0(\varphi) = \chi(\varphi)_{H \to 0} = dm(H,\varphi)/dH|_{H \to 0}$ at $T = 150$ K for the sample frozen at $H_{cool} = 5$ kOe. Its angular dependence is described by expression

$$\chi_0(\varphi) = \chi_0^{(\perp)} + (\chi_0^{(\|)} - \chi_0^{(\perp)})\cos^2 \varphi , \qquad (1)$$

where $\chi_0^{(\perp)}$ is the susceptibility in $\mathbf{H} \perp \mathbf{H}_{cool}$, and $\chi_0^{(\|)}$ is the susceptibility in $\mathbf{H} \| \mathbf{H}_{cool}$. It follows from the angular dependence (1) that the given anisotropy of the sample at $T = 150$ K is uniaxial anisotropy of the second order.

## 6. Temperature dependences of magnetization

The occurrence of anisotropy is associated with displacements (and/or reorientations) of the particles at room temperature. Below a particular temperature (in our case, approximately 220 K), the mobility of inclusions is blocked by a rigidized matrix. With a further decrease of temperature, the ensemble of particles must preserve the anisotropy of the arrangement of the particles (microstructure of the filler), which they acquired in a magnetic field at high temperature. Conversely, at a temperature above the "blocking temperature", the particles are unblocked and they become capable of moving in the matrix when magnetized. Thus, unlike conventional magnetic nanocomposites, where the blocking effect of the directions of the magnetic moments of particles by their magnetic anisotropy is observed[65-67], the effect of blocking of the mobility of particles should be detectable in the MAE when the temperature is lowered.

The effect of the described immobilization ("blocking") of inclusions on the magnetic properties of the MAE is clearly seen in the temperature dependence of the magnetization obtained in ZFC (zero-field-cooled) measurements. Fig. 5 shows the temperature behavior of the sample frozen in the absence of magnetic field ($H_{cool} = 0$). The sample was placed in a magnetic field $H = 1.5$ kOe ($\mathbf{H} \perp OZ$) at $T = 150$ K. Then, keeping the external field constant, the sample was heated up to the room temperature and its magnetization $m$ was measured simultaneously. In the temperature range between 220 K and 230 K there is a significant increase of the magnetization which can be explained by softening of the matrix. For $T < 220$ K, the positions of the particles are blocked, there is a slight decrease in the magnetization of the sample with the increasing temperature. When the matrix becomes compliant, the easy-axis anisotropy is induced along $\mathbf{H}$, and the magnetization significantly increases with growing temperature as a result of the appearance of this anisotropy.

Although these measurements were aimed at revealing the appearance of anisotropy induced by the magnetic field, it turned out that the softening of the matrix is also accompanied by the appearance of hysteresis in the magnetization reversal curves. At the beginning of the transition temperature range ($T = 220$ K) there is practically no hysteresis of the magnetization, and at the end the transition range ($T = 230$ K) the observed hysteresis does not differ significantly from the loop shown in Fig. 1.

Figure 6 also displays the temperature dependence of the magnetization obtained with FCW (field cooled warming) measurements. The sample was first cooled in $H_{cool} \perp OZ$ with $H_{cool} = 4$ kOe and the temperature dependence of magnetization has been measured upon heating in the field $H = 1.5$ kOe ($H \| H_{cool}$), which is smaller than the cooling field. A steep decrease of the magnetization is observed the temperature interval between 220 and 230 K, which is also related to the unblocking of the filler particles (softening of the matrix). At $T < 220$ K, the sample has a magnetization greater than at $T > 230$ K, because the magnetic anisotropy of the sample cooled in $H_{cool} = 4$ kOe has the larger magnetic anisotropy, than the anisotropy acquired in the field $H = 1.5$ kOe after unblocking of particles.

Figure 6 also presents the temperature dependence of magnetization obtained with FC (field-cooled) measurements. The magnetization was measured upon cooling of the sample in a magnetic field $H_{cool} = 4$ kOe. This dependence has no noticeable changes in the temperature range between 220 K and 230 K. The explanation can be that the blocking of particles does not modify the already formed microstructure of the filler and the anisotropy remains unchanged.

We note that the curves in Fig. 6 do not coincide at room temperature. This is due to the fact that at room temperature the magnetization in the field $H = 1.5$ kOe falls into the hysteresis region (cf. Fig. 2). However, we cannot use smaller measurement fields, because the anisotropy will become weaker.

Thus, from the discussed dependencies it follows that the magnitude of the anisotropy induced by the magnetic field depends on the magnitude of the applied field. It is found that the magnetic anisotropy induced at room temperature in the MAE leads to the peculiar behavior of the temperature dependences of the magnetization in the ZFC, FC, and FCW experiments, when the matrix is solidified and blocks the particle displacements.

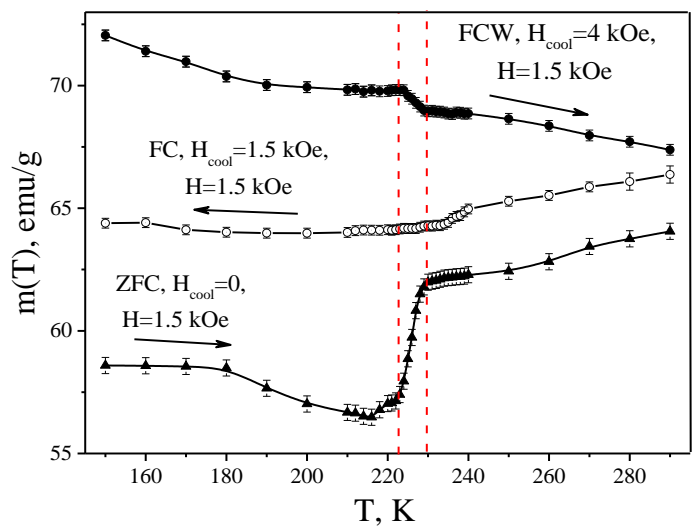

**Figure 6.** Temperature dependences of magnetization for FC, ZFC and FCW measurements. The direction of temperature variation is indicated by arrows. The transition temperature range of the MAE matrix is outlined by dashed vertical lines.

## 7. Magnetic anisotropy energy

In Section 5 above, it was experimentally shown that the frozen sample has the uniaxial magnetic anisotropy. In this case, the magnetic anisotropy energy density $E_A$ should have the form of

$$E_A(\varphi) = -\frac{1}{2} K_{ef}(H_{cool}) \cos^2 \varphi, \qquad (2)$$

where the effective anisotropy constant $K_{ef}$ is a function of the field $H_{cool}$.

Formula (2) can be easily verified experimentally. The anisotropy constant is $K_{ef}(H_{cool}) = 2 \cdot (E_A(\varphi = 90°) - E_A(\varphi = 0°)) = 2\Delta E_A(H_{cool})$, where $\Delta E_A$ is the peak-to-peak amplitude of (2). The latter is equal to the integral between the two curves $m(H)$ for the states when $\boldsymbol{H} \| \boldsymbol{H}_{cool}$ and $\boldsymbol{H} \perp \boldsymbol{H}_{cool}$ (cf. Fig. 3).

$$\begin{aligned}\Delta E_A(H_{cool}) &= \int_0^\infty (H_{int}(\perp H_{cool})dm - H_{int}(\| H_{cool})dm) = \\ &= \int_0^\infty (m(H_{int} \| H_{cool}) - m(H_{int} \perp H_{cool}))dH_{int} = \\ &= \int_0^\infty (m(H \| H_{cool}) - m(H \perp H_{cool}))dH. \end{aligned} \qquad (3)$$

The equality between the last two integrals in (3) is satisfied if the components of the demagnetizing factor are the same for both orientations of magnetizations, which, as it can be seen from sections 3 and 4 of this paper, is satisfied for the case when the involved magnetic fields are in the plane of the sample, $\boldsymbol{H}_{cool}, \boldsymbol{H} \perp OZ$. Obviously, calculation (3) can be performed with fixed values $\varphi = 0°$ and $90°$ for different values of $H_{cool}$ and with $H_{cool} = const$ for different angles $\varphi$ (see Fig. 5).

To characterize the magnetic anisotropy, the magnetic anisotropy field $H_A$ is often convenient. Its value is directly proportional to the energy of the magnetic anisotropy:

$$H_A(H_{cool}) = 2\Delta E_A(H_{cool}) / m_{max}, \qquad (4)$$

where $m_{max}$ is the maximum magnetization of the sample at $H \to \infty$.

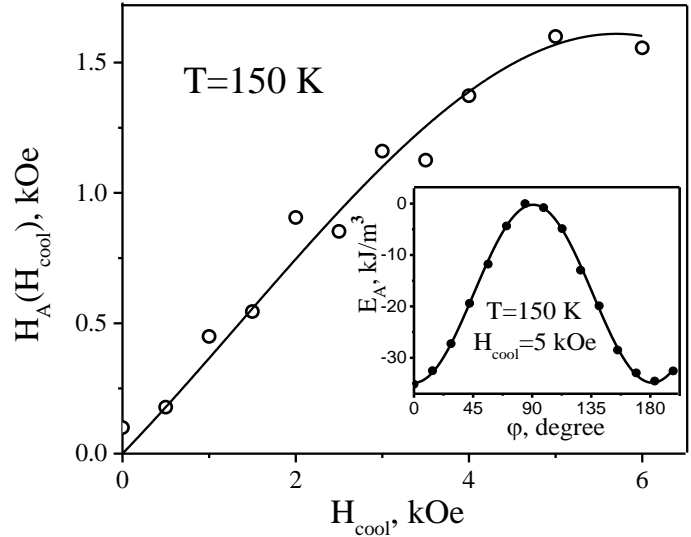

**Figure 7.** Dependence of the induced anisotropy field $H_A(H_{cool})$ on the magnitude of the magnetic field $H_{cool}$, in which the sample was cooled down to $T = 150$ K. The inset shows the dependence for the anisotropy energy density $E_A(\varphi)$ measured for $H_{cool} = 5\,\text{kOe}$.

The value of $H_A$ must depend on the magnitude of the field $H_{cool}$ that induced it. From the measurements of the low-temperature field dependences of the magnetization of the sample frozen in different $H_{cool}$, we obtained, using expressions (3) and (4) the dependence of the anisotropy field $H_A$ on $H_{cool}$, shown in Fig. 7. The experimental points are designated by open circles, and the solid curve is a fit represented by an expression $H_A(H_{cool}) = a_1 H_{cool} + a_2 H_{cool}^2 + a_3 H_{cool}^3$ with coefficients $a_1 = 0.35$, $a_2 = 2.75 \cdot 10^{-2}$ (kOe)$^{-1}$, $a_3 = -6.75 \cdot 10^{-3}$ (kOe)$^{-2}$. This polynomial expression is shown as a guide to the eye. It is seen that $H_A$ grows with increasing $H_{cool}$ and there is some indication of saturation for $H_{cool} > 5$ kOe.

Using the dependences of the magnetization obtained in $H$, inclined with respect to $H_{cool}$ as in Fig. 5, it is possible to obtain the dependence of the magnetic anisotropy energy density on the inclination angle $\varphi$ of the field $H$ in the state with saturated magnetization. The resulting dependence of $E_A$ for $T = 150$ K and $H_{cool} = 5\,\text{kOe}$ is shown on the inset in Fig. 7. The experimental data on the insert are indicated by filled circles, and the solid line denotes the result of fitting the dependence by the expression (2) with $K_{ef} = 6.9 \cdot 10^4\,\text{J/m}^3$. The validity of expression (2), in addition to (1), confirms the uniaxial second-order type of magnetic anisotropy induced in the MAE by the magnetic field.

## 8. Discussion

It is shown that in an MAE sample with fixed dimensions, although it is not a magnetically homogeneous material and the ferromagnetic filler particles can be displaced and/or reoriented within the sample, the influence of the shape anisotropy of the sample can be described by means of the demagnetizing factor, as in magnetically homogeneous ferromagnetic materials, and the components of this factor do not depend on the magnetic field and temperature.

We also obtained that the additional contribution to the inter-particle interaction from the mutual arrangement of magnetic particles in the MAE can be described as magnetic anisotropy induced by an external magnetic field and associated with the change of the mutual positions of the magnetized particles under the influence of magnetic fields. The value of this anisotropy depends on the magnitude of the external magnetic field applied to the sample when the matrix is compliant. Measurements at $T = 150$ K show that the larger is the external magnetic field induced this anisotropy, the greater is the magnetic anisotropy energy density. This means that in an MAE sample, when it is magnetized at a temperature above the blocking temperature for the particle displacements, a magnetic anisotropy is induced, which grows with the increasing external magnetic field. For the saturating magnetic field in the MAE at room temperature, the magnitude of the magnetic anisotropy can be estimated from Fig. 7 at $H_{cool} = 6$ kOe. Thus, when the MAE sample of interest is magnetized to saturation at room temperature, the maximum anisotropy field is approximately equal to 1.6 kOe, and the effective magnetic anisotropy constant of the sample is equal to $K_{ef} = 6.9 \cdot 10^4$ J/m$^3$, which is the effective constant of the MAE sample as a whole. This value of the magnetic anisotropy constant is almost twice as large as the shear modulus of this MAE in the absence of an external magnetic field (see Ref. 14, where the low-frequency magnetorheological properties of this MAE were reported).

The effective anisotropy constant of soft magnetic filler particles in the MAE can be estimated as

$$\tilde{K}_{ef} = K_{ef} / \varphi_V \approx 3 \cdot 10^5 \text{ J/m}^3, \qquad (5)$$

where $\tilde{K}_{ef}$ is the effective magnetic anisotropy constant, re-calculated as the energy density per volume occupied by the ferromagnetic inclusions in the MAE sample in the saturating field. As seen from (5), the magnetic anisotropy constant of the particles of the MAE under investigation is almost an order of magnitude greater than its effective low-frequency shear storage modulus $G'_0 \approx 40$ kPa[14]. The PDMS matrix is even softer, its low-frequency shear storage modulus $\mu$ is about 7 kPa[29]. Therefore, the ratio of $\tilde{K}_{ef} / \mu$ is approximately 40. It has been recently shown by us that large ratio $\tilde{K}_{ef} / \mu$ may lead to large magnetorheological effects in highly filled MAEs through the renormalization of the critical index or the percolation threshold (as first suggested in Ref. 68) due to single-particle magnetostriction mechanism[69,70]. Note that in conventional magnetic composite materials with fixed filler particles, the percolation threshold does not depend on external magnetic field[71]. Thus, we propose that induced magnetic anisotropy requires attention for the explanation of the magnetorheological properties of MAE. It may lead to additional rigidity of the elastic subsystem by at least the value of the effective magnetic anisotropy constant, increasing the shear modulus of the composite material by its value.

If MAE particles have an elongated shape in the form of ellipsoids of revolution, then under the action of the magnetic field they will rotate, trying to arrange their long axes along the field. For particles with a ratio of the length of the larger axis of the ellipsoid to the length of its smaller axis approximately equal to 1.2[70], the magnitude of their anisotropy field will be comparable to that obtained by us in the experiment. In this case it is required that in a strong field all the particles in their large axes are oriented along the field. The possibility of such rotations is discussed in the literature. Such an anisotropy of the MAE makes single-particle magnetostriction mechanism[11,70], which is connected with the rotation of particles in a magnetic field, relevant.

However, it is seen in Figure 4 that (although the curve 3 ($H_{cool} = 0$) goes, as it should be expected, between the curves 1 ($\boldsymbol{H} \| \boldsymbol{H}_{cool}$) and 2 ($\boldsymbol{H} \perp \boldsymbol{H}_{cool}$), where $H_{cool} \neq 0$) the curve 3 is shifted noticeably towards the curve 2. Therefore, the interpretation of the curve set 1 – 3 cannot be done in the framework of the independent-grain approximation[72] where the sequence of curves corresponds to the cases of the magnetic field applied along the long ellipsoid axes, randomly oriented ellipsoids and the magnetic field directed perpendicular to the long axes, respectively. It is essential to consider the effect of particles displacement on the susceptibility of the MAE sample.

The anisotropy may be also created by spherical particles due to the ubiquitous presence of a substantial number of multi-particle clusters in real MAEs; a number of electron microscopy evidences can be found in the literature[73-75].

Phenomenologically, the energy density of the uniaxial magnetic anisotropy of the frozen MAE, induced by the magnetic field $H_{cool}$, can be written in the following form:

$$E_A(H_{cool}) = -\frac{1}{2} K_{ef}(H_{cool}) \frac{m_H^2}{m_{max}^2}, \tag{6}$$

where $K_{ef}(H_{cool})$ is value of the anisotropy constant of the frozen MAE, which is dependent on the field $H_{cool}$. $m_H$ is the projection of the MAE magnetization on the vector $\boldsymbol{H}$, if $\boldsymbol{H}$ is not collinear with $\boldsymbol{H}_{cool}$. It is $m_H$ that is experimentally measured in an inclined field. For $m_H = m_{max}$ and $\boldsymbol{H}$ inclined by an angle $\varphi$ with respect to $\boldsymbol{H}_{cool}$, the formula (6) becomes identical to expression (2).

However, we note that the magnetic anisotropy does not arise from the cooling of the sample. It occurs when the sample is magnetized at room temperature, when the matrix is soft elastic. This means that at room temperature the energy of the MAE magnetized by the field $H$ will be reduced by the addition of

$$\Delta E = -\frac{1}{2} K(H) \frac{m^2(H)}{m_{max}^2}, \tag{7}$$

where $K(H) = K(H_{cool} = H)$, $m(H)$ is the magnetization of the MAE in the field $H$, $\boldsymbol{m} \| \boldsymbol{H}$. The constant $K(H)$, as it follows from the experiment, depends linearly on $H$ for small (< 3kOe) fields and shows a trend to the saturation in strong (> 4 kOe) fields, that is how the MAE differs from composite materials with immobilized particles.

It follows from (7) that at room temperature, in addition to the external field **H**, the effective field $H_{ef}$ is also forceful in the MAE sample:

$$H_{ef}(H) = K(H)\frac{m(H)}{m_{max}^2}, \qquad (8)$$

and besides $H_{ef}(H \to \infty) = H_A(H_{cool} \to \infty)$.

As can be seen from (8), the magnetization of MAE is characterized by an unusual type of nonlinearity that is absent both in composite materials with fixed filler particles[76] and in conventional magnetic materials. The difference of the present work to conventional materials (as, e.g., in Ref. 76) is that it is possible to control the magnitude of the magnetic anisotropy by external magnetic field and fix it at low temperature.

We emphasize that in this work, by measuring the anisotropy field of a frozen sample, we have experimentally found the effective field acting on the magnetization when MAE is magnetized at room temperature. The elucidation of the origin of the appearance of an additional field in a magnetized MAE is one of the main challenges, the solution of which will make it possible to understand the characteristics of magnetism in MAEs.

Let us make the main conclusion of this work. In MAEs, when magnetized, the particles are displaced by the influence of magnetic forces, which leads to the formation of an additional magnetic field directed along the external field and to the uniaxial magnetic anisotropy, which contributes to the easier magnetization of the MAE with the larger susceptibility. The formation of magnetic anisotropy and the corresponding characteristic field is a consequence of the nonlinear, self-consistent MAE magnetization process: the larger the field, the higher the magnetization, and the greater the particle displacement, which leads to an increase in the magnetic anisotropy, and, ultimately, to an increase of the magnetization.

We have experimentally observed the magnetic anisotropy in a compliant MAE, filled with soft magnetic particles, induced by external magnetic fields. The readout and measurement of this anisotropy is possible at low temperatures when the elastomer matrix becomes rigid. The observed anisotropy is uniaxial and its magnitude grows with increasing magnetic field. The energy density of magnetic anisotropy can significantly exceed the efficient shear modulus of the composite materials. We believe that induced magnetic anisotropy should be taken into account for explaining the large magnetorheological and magnetic stiffening effects in MAE materials. In this context, the reported results and experimental methodology can be useful for developing MAE materials for such target applications as vibration absorbers and isolators[1,77].

## Acknowledgements

V.M.K, A.A.S and M.S. thank the Bayerische Forschungsallianz for financial support of the reciprocal visits (grant No. BayIntAn_OTHR_2017_120). The work of M.S. was funded by the Deutsche Forschungsgemeinschaft (DFG, German Research Foundation) – project number 389008375 and OTH Regensburg (internal cluster funding).

**References**

1. Y. Li, J. Li, W.H. Li, and H. Du, Smart Mater. Struct. **23**, 123001 (2014).

2. Ubaidillah, J. Sutrisno, A. Purwanto, and S. A. Mazlan, Adv. Eng. Mater. **17**, 563 (2015).

3. A. M. Menzel, Phys. Reps. **554**, 1 (2015).

4. M.A. Cantera, M. Behrooz, R.F. Gibson, and F. Gordaninejad, Smart Mater. Struct. **26**, 023001 (2017).

5. L.A. Makarova, V.V. Rodionova, Y.A. Alekhina, T.S. Rusakova, A.S. Omelyanchik, and N.S. Perov, IEEE Trans. Magn. **53**, 1 (2017).

6. R. Weeber, R., M. Hermes, A.M. Schmidt, and C. Holm, J. Phys.: Cond. Matt. **30**(6), 063002 (2018).

7. G.V. Stepanov, D.Yu. Borin, Yu.L. Raikher, P.V. Melenev, and N.S. Perov, J. Phys.: Cond. Matter **20**, 204121 (2008).

8. S. Abramchuk, E. Kramarenko, G. Stepanov, L.V. Nikitin, G. Filipcsei, A.R. Khokhlov, and M. Zrinyi, Polym. Adv. Technol. **18**, 883 (2007).

9. H.-N. An, S. J. Picken, and E. Mendes, Soft Matter **8**, 11995 (2012).

10. M. Schümann and S. Odenbach, J. Magn. Magn. Mater. **441**, 88 (2017).

11. V.M. Kalita, A.A. Snarskii, M. Shamonin, and D. Zorinets, Phys. Rev. E **95**, 032503 (2017).

12. P.A. Sánchez, T. Gundermann, A. Dobroserdova, S.S. Kantorovich, and S. Odenbach, Soft Matter **14**(11), 2170-2183 (2018).

13. I.A. Belyaeva, E.Y. Kramarenko, G.V. Stepanov, V.V. Sorokin, D. Stadler, and M. Shamonin, Soft Matter **12**(11), 2901 (2016).

14. A.V. Bodnaruk, A. Brunhuber, V.M. Kalita, M.M. Kulyk, A.A. Snarskii, A. F. Lozenko, S.M. Ryabchenko, and M. Shamonin, J. Appl. Phys. **123**(11), 115118 (2018).

15. N. Bosq, N. Guigo, J. Persello, and N. Sbirrazzuoli, Phys. Chem. Chem. Phys., **16**, 7830 (2014).

16. J. Ginder, S. Clark, W. Schlotter, and M. Nichols, Int. J. Modern Phys. B **16**, 2412 (2002).

17. L. Nikitin, G. Stepanov, L. Mironova, and A. Gorbunov, J. Magn. Magn. Mater. **272**, 2072 (2004).

18. E. Coquelle and G. Bossis, J. Adv. Sci. **17**, 132 (2005).

19. V. Q. Nguyen, A. S. Ahmed, and R. V. Ramanujan, Adv. Mater. **24**, 4041 (2012).

20. X. Gong, G. Liao, and S. Xuan, Appl. Phys. Lett., **100**, 211909 (2012).

21. G. Stepanov, E. Y. Kramarenko, and D. Semerenko, J. Phys.: Conf. Ser., 012031 (2013).

22. M. Lopez-Lopez, J. D. Durán, L. Y. Iskakova, and A. Y. Zubarev, J. Nanofluids **5**, 479 (2016).

23. S. Odenbach, Arch. Appl. Mech. **86**, 269 (2016).

24. A. Chertovich, G. Stepanov, E. Y. Kramarenko, and A. Khokhlov, Macromol. Mater. Eng. **295**, 336 (2010).

25. A. Stoll, M. Mayer, G. J. Monkman, and M. Shamonin, J. Appl. Polym. Sci. **131**, 39793 (2014).

26. J. Yao, Y. Sun, Y. Wang, Q. Fu, Z. Xiong, and Y. Liu, Compos. Science Technol. **162**, 170 (2018).

27. I. Bica, Y. D. Liu, and H. J. Choi, Colloid Polym. Sci. **290**, 1115 (2012).

28. A.S. Semisalova, N.S. Perov, G.V. Stepanov, E.Y. Kramarenko, and A.R. Khokhlov, Soft Matter **9**, 11318 (2013).

29. I.A. Belyaeva, E.Y. Kramarenko, and M. Shamonin, Polymer **127**, 119 (2017).

30. I. Bica, Mater. Lett. 63, 2230 (2009).

31. N. Kchit, P. Lancon, and G. Bossis, J. Phys. D: Appl. Phys. **42**, 105506 (2009).

32. G. V Stepanov, D. A. Semerenko, A. V Bakhtiiarov, and P. A. Storozhenko, J. Supercond. Nov. Magn. **26**, 1055 (2013).

33. M. Yu, B. Ju, J. Fu, S. Liu, and S.-B. Choi, Ind. & Eng. Chem. Res. **53**, 4704 (2014).

34. G. Ausanio, V. Iannotti, E. Ricciardi, L. Lanotte, and L. Lanotte, Sens. Actuat. A: Phys. **205**, 235 (2014).

35. Y. Wang, S. Xuan, L. Ge, Q. Wen, and X. Gong, Smart Mater. Struct. **26**, 015004 (2016).

36. M. Schümann, J. Morich, T. Kaufhold, V. Böhm, K. Zimmermann, and S. Odenbach, J. Magn. Magn. Mater. **453**, 198 (2018).

37. J. De Vicente, G. Bossis, S. Lacis, and M. Guyot, J. Magn. Magn. Mater. **251**, 100 (2002).

38. G.V. Stepanov, D.Y. Borin, S. Odenbach, and A.I. Gorbunov, Solid State Phenom. **152**, 190 (2009).

39. J. Zeng, Y. Guo, Y. Li, J. Zhu, and J. Li, J. Appl. Phys. **113**, 17A919 (2013).

40. A. Yu. Zubarev, D. N. Chirikov, D. Yu. Borin, and G. V. Stepanov, Soft Matter **12**, 6473 (2016).

41. D.Y. Borin and S Odenbach, J. Magn. Magn. Mater. **431**, 115 (2017).

42. M. Krautz, D. Werner, M. Schrödner, A. Funk, A. Jantz, J. Popp, J. Eckert, and A. Waske, J. Magn. Magn. Mater. **426**, 60 (2017).

43. B. Xu, C. Paillard, B. Dkhil, and L. Bellaiche, Phys. Rev. B **94**(14), 140101 (2016).

44. Z. Biolek, and D. Biolek, IEEE Trans. Circ. Syst. II: Expr. Briefs **61**(7), 491 (2014).

45. A.O. Ivanov, Z. Wang, and C. Holm, Phys. Rev. E **69**, 031206 (2004).

46. A.M. Biller, O.V. Stolbov, and Y.L. Raikher, Phys. Rev. E **92,** 023202 (2015).

47. D. Romeis, V. Toshchevikov, and M. Saphiannikova, Soft Matter **12**(46), 9364-9376 (2016).

48. G. Diguet, E. Beaugnon, and J.Y. Cavaillé, J. Magn. Magn. Mater. **322**(21), 3337 (2010).

49. C. Kittel, *Introduction to solid state physics* 8th Edition, New York: Wiley (2005).

50. H. An, S.J. Picken, and E. Mendes, Polymer **53**(19), 4164-4170 (2012).

51. A. Butera, N. Álvarez, G. Jorge, M.M. Ruiz, J.L. Mietta, and R.M. Negri, Phys. Rev. B **86**(14), 144424 (2012).

52. T. Tian, and M. Nakano, J. Intell. Mater. Syst. Struct. **29**(2), 151-159 (2018).

53. D. Ivaneyko, V. Toshchevikov, and M. Saphiannikova, Polymer, **147**, 95-107 (2018).

54. O.V. Stolbov, Y.L. Raikher, G.V. Stepanov, A.V. Chertovich, E.Y. Kramarenko, and A.R. Khokhlov, Polymer Sc. Ser. A **52**(12), 1344 (2010).

55. I. Agirre-Olabide, A. Lion, and M.J. Elejabarrieta, Smart. Mater. Struct. **26**, 035021 (2017).

56. G. Filipcsei, I. Csetneki, A. Szilágyi, and M. Zrínyi, In *Oligomers-Polymer Composites-Molecular Imprinting*, 137, Springer, Berlin, Heidelberg (2007).

57. M.A. Abshinova, I. Kuřitka, N.E. Kazantseva, J. Vilčáková, and P. Sáha, Mater. Chem. Phys. **114**(1), 78 (2009).

58. M. Schrödner, and G. Pflug, J. Magn. Magn. Mater. **454**, 258 (2018).

59. J.M. Coey, *Magnetism and magnetic materials*. Cambridge University Press, Cambridge, UK (2010).

60. LM. Palacios-Pineda, I.A. Perales-Martinez, L.M. Lozano-Sanchez, O. Martínez-Romero, J. Puente-Córdova, E. Segura-Cárdenas, and A. Elías-Zúñiga, Polymers, **9**(12), 696 (2017).

61. D. Ivaneyko, V. Toshchevikov, M. Saphiannikova, and G. Heinrich, Cond. Matter Phys. **15**, 33601 (2012).

62. S.V. Kredentser, M.M. Kulyk, V.M. Kalita, K.Y. Slyusarenko, V.Y. Reshetnyak and Y.A. Reznikov, Soft Matter **13**(22), 4080 (2017).

63. F. Martinez-Pedrero, A. Cebers, and P. Tierno, Soft Matter **12**(16), 3688 (2016).

64. F. Martinez-Pedrero, A. Cebers, and P. Tierno, Phys. Rev. Appl. **6**(3), 034002 (2016).

65. V. Markovich, I. Fita, A. Wisniewski, D. Mogilyansky, R. Puzniak, L. Titelman, C. Martin and G. Gorodetsky, Phys. Rev. B **81**(9), 094428 (2010).

66. F.H. Sánchez, P. Mendoza Zélis, M.L. Arciniegas, G.A. Pasquevich, and M.F. van Raap, Phys. Rev. B **95**(13), 134421 (2017).

67. V.M. Kalita, D.M. Polishchuk, D.G. Kovalchuk, A.V. Bodnaruk, S.O. Solopan, A.I. Tovstolytkin, S.M. Ryabchenko and A.G. Belous, Phys. Chem. Chem. Phys. **19**(39), 27015 (2017).

68. T. Mitsumata, S. Ohori, A. Honda, and M. Kawai, Soft Matter **9**(3), 904 (2012).

69. A.A. Snarskii, V.M. Kalita, and M. Shamonin, Scientific Reports **8**(1), 4397 (2018).

70. V.M. Kalita, A.A. Snarskii, D. Zorinets, and M. Shamonin, Phys. Rev. E **93**(6), 062503 (2016).

71. T.J. Fiske, H.S. Gokturk, and D.M. Kalyon, J. Mater. Sci. **32**(20), 5551 (1997).

72. A.G. Gurevich, and G.A. Melkov, *Magnetization oscillations and waves*, CRC Press (1996).

73. O.V. Stolbov, Y.L. Raikher, and M. Balasoiu, Soft Matter **7**(18), 8484-8487 (2011).

74. G.V. Stepanov, S.S. Abramchuk, D.A. Grishin, L.V. Nikitin, E.Y. Kramarenko, and A.R. Khokhlov, Polymer **48**(2), 488-495 (2007).

75. K. Shimada, S. Shuchi, and H. Kanno, J. Intell. Mater. Syst. Struct. **16**(1), 15 (2005).

76. M. Anhalt, B. Weidenfeller, J. Appl. Phys. **101**(2), 023907 (2007).

77. L.J. Qian, F.L. Xin, X.X. Bai, and N.M. Wereley, J. Intell. Mater. Syst. Struct. **28**(18), 2539 (2017).